%% file: bf_etaprime_0319.tex
\documentclass[twocolumn,showpacs,aps,prl,amsmath,amssymb]{revtex4-1}
\usepackage[dvips]{graphicx}
\usepackage{subfigure}
\usepackage{color}
\usepackage{epsfig}
\usepackage{indentfirst}
\usepackage{multirow}
\usepackage{amsmath}
\usepackage{amssymb}
\usepackage{bm}
\usepackage{fancyhdr}
\usepackage{enumerate}
\usepackage[perpage,symbol]{footmisc}
\usepackage[bookmarksnumbered,pdfstartview=FitH,colorlinks=true,menucolor=black,linkcolor=blue]{hyperref}
\usepackage{xspace}
\definecolor{orange}{RGB}{255,165,0}
\definecolor{chocolate}{RGB}{210,105,30}
\definecolor{purple}{RGB}{255,0,255}
\usepackage{epstopdf}
\usepackage{array}

\newcommand{\jpsi}{J/\psi}
\newcommand{\pip}{\pi^+}
\newcommand{\pim}{\pi^-}
\newcommand{\pio}{\pi^{0}}
\newcommand{\ra}{\rightarrow}

\newcommand{\epem}{e^+e^-}
\newcommand{\etap}{\eta^{\prime}}

\begin{document}

\title{\boldmath Precision measurement of the branching fractions of $\etap$ decays}

\input{authors.tex}

\date{\today}


\begin{abstract}

Based on a sample of $(1310.6 \pm 7.0) \times 10^6~\jpsi$ events collected with the BESIII detector, we present
measurements of $J/\psi$ and $\eta^\prime$ absolute branching fractions using the process $J/\psi\rightarrow\gamma\eta^\prime$.
By analyzing events where the radiative photon converts into an $e^+e^-$ pair, the branching fraction for $\jpsi \ra \gamma \etap$
is measured to be $(5.27\pm0.03\pm0.05)\times 10^{-3}$.
The absolute branching fractions of the five dominant decay channels of the $\eta^\prime$ are then measured for the first time
and are determined to be
$\mathcal{B}(\etap \ra \gamma \pi^{+} \pi^{-})$ = (29.90$\pm$0.03$\pm$0.55)\%,
$\mathcal{B}(\etap \ra \eta \pi^{+} \pi^{-})$ = (41.24$\pm$0.08$\pm$1.24)\%,
$\mathcal{B}(\etap \ra \eta \pi^{0} \pi^{0})$ = (21.36$\pm$0.10$\pm$0.92)\%,
$\mathcal{B}(\etap \ra \gamma \omega)$ = (2.489$\pm$0.018$\pm$0.074)\%,
and $\mathcal{B}(\etap \ra \gamma \gamma)$ = (2.331$\pm$0.012$\pm$0.035)\%,
where the first uncertainties are statistical and the second systematic.

\end{abstract}


\maketitle


Even though the main properties of the $\eta^\prime$ meson are
firmly established and its main decay modes are fairly well known, it still attracts both theoretical and experimental attention due to its special role in understanding low energy Quantum Chromodynamics~(QCD).
Decays of the $\eta^\prime$ meson have inspired the study of a wide variety of physics issues,  $e.g.$ $\eta-\eta^\prime$ mixing, the light quark masses, as well as physics
beyond the Standard Model.  Hence considerable theoretical effort has been devoted to investigate its decay dynamics and partial decay widths with different approaches~\cite{th1,th2,th3,th4,th5,th6}.
However, no absolute branching fractions~(BFs) of $\eta^\prime$ decays have yet been measured due to the difficulty of tagging its inclusive decays.  The exclusive BFs of the $\eta^\prime$ summarized by
the Particle Data Group~(PDG)~\cite{Olive:2016xmw} are all relative measurements.
The two most precise measurements so far are from the BES and CLEO experiments. The BES experiment~\cite{Ablikim:2005je}
reported the relative BFs of $\mathcal{B}(\etap \ra \gamma \gamma)/\mathcal{B}(\etap \ra \gamma \pip \pim)$ and $\mathcal{B}(\etap \ra \eta \pip \pim)/\mathcal{B}(\etap \ra \gamma \pip \pim)$, while the CLEO experiment \cite{Pedlar:2009aa} measured the
branching fractions of its five decay modes by constraining their sum to be $(99.2\pm0.2)\%$.
The absolute BF measurement of the five dominant decay modes are also essential in order to improve the precision of the BFs for several $\etap$ decays, which are obtained via normalization to the dominant $\etap$ decay modes.

In this Letter, we develop an approach to measure the absolute BFs of the exclusive decays of the $\eta^\prime$ meson using a sample of $(1310.6 \pm 7.0) \times 10^6$ $J/\psi$ events~\cite{Ablikim:2016fal} collected with the BESIII detector.
The design and performance of the BESIII detector are described in detail in Ref.~\cite{Ablikim:2009aa}.
Taking advantage of the excellent momentum resolution of charged tracks in the Main Drift Chamber~(MDC), photon conversions to $\epem$ pairs provide a unique tool to reconstruct the inclusive photon spectrum from radiative $J/\psi$ decays. Take $J/\psi\rightarrow\gamma\eta^\prime$ for example, Monte Carlo~(MC) study indicates that the energy resolution of the radiative photon could be improved by a factor of three using the photon conversion events.
This enables us to tag the $\eta^\prime$ inclusive decays and then to measure the absolute BF of $J/\psi\rightarrow\gamma\eta^\prime$, using
\begin{equation}
    \mathcal{B}(\jpsi \to \gamma \etap) = \frac{{{N^{\rm obs}_{J/\psi\rightarrow\gamma\eta^\prime}}}}{{N_{\jpsi}\cdot \varepsilon\cdot f}},
\label{eq:gametapbf}
\end{equation}	
where $N^{\rm obs}_{J/\psi\rightarrow\gamma\eta^\prime}$ is the observed $\etap$ yield, $\varepsilon$ is the detection efficiency obtained from MC simulation, and $N_{\jpsi }$ is the  number of $\jpsi$ events. The photon conversion process is simulated with GEANT4~\cite{Agostinelli:2002hh}, and $f$ is a correction factor to account for the difference in the photon conversion efficiencies between data and MC simulation.

After the $\etap$ inclusive measurement, we present precision measurements of $\eta^\prime$ decays to $\gamma\pi^+\pi^-$, $\eta\pi^+\pi^-$, $\eta\pi^0\pi^0$, $\gamma\omega$ and $\gamma\gamma$, again using $\jpsi$ decays to $\gamma \etap$, but with the radiative photon directly detected by the Electromagnetic Calorimeter~(EMC) to improve the statistics.
With the help of Eq.~(\ref{eq:gametapbf}),  the BF for each $\eta^\prime$ exclusive decay is then calculated using
\begin{equation}
	\mathcal{B}(\etap \to X) = \frac{{{N_{\etap \to X }^{\rm obs}}}}{{{\varepsilon _{\etap \to X }}}} \cdot \frac{{{\varepsilon}}}{{{N^{\rm obs}_{J/\psi\rightarrow\gamma\eta^\prime}}}} \cdot f,
\label{eq:gamgambf}
\end{equation}
where $N_{\etap \to X}^{\rm obs}$ is the number of signal events obtained from a fit to data and $\varepsilon_{\etap \to X }$ is the MC-determined reconstruction efficiency.


For the process $\jpsi \ra \gamma \etap$ where the radiative photon converts to an $e^+e^-$ pair,
candidate events are required to have at least two oppositely charged tracks.
Each charged track is reconstructed using information from the MDC and is required to have a polar angle in the range $|\cos{\theta}| < 0.93$ and pass within $\pm30$~cm
of the
interaction point along the beam direction.
To reconstruct the photon conversions, a photon conversion finder ~\cite{Xu:2012xq} is applied to
all combinations of track pairs with opposite charge.
The photon conversion point~(CP) is reconstructed using the two charged track trajectories in
the $x$-$y$ plane, which is perpendicular to the beam line.
The photon conversion length $R_{{xy}}$ is defined as the distance
from the beam line to the CP in the $x$-$y$ plane.
Photon conversion events accumulate at $R_{xy}=3$~cm and $R_{xy}=6$~cm corresponding to the position of
the beam pipe and the inner wall of the the MDC. The detail studies illustrate that
the distributions of $R_{xy}$ for data and MC simulations are consistent with each other, as presented in Ref.~\cite{Xu:2012xq}.

To reduce the large combinatorial background from $\pi^0 \ra \gamma\gamma$ decays where one of the photons converts into an $e^+e^-$ pair, the $e^+e^-$ pairs that, when combined with a photon candidate, form a $\pi^0$ candidate with an invariant mass within 20 MeV/$c^2$ of the $\pi^0$ mass~(corresponding to $\pm 3$ times the mass resolution) are not used in the reconstruction.
Candidate events with one photon depositing more than 1.2~GeV in the EMC are rejected to suppress background from $e^+e^-\rightarrow\gamma\gamma(\gamma)$.
A MC study demonstrated that a peaking background contribution is from the electromagnetic Dalitz decay \cite{Landsberg:1986fd} $\jpsi \to \etap \epem$, which can be effectively removed by
requiring $R_{xy}~>$ 2~cm.

\begin{figure}[htbp]
	\centering
	\begin{subfigure}
	{
		\includegraphics[width = 0.22\textwidth]{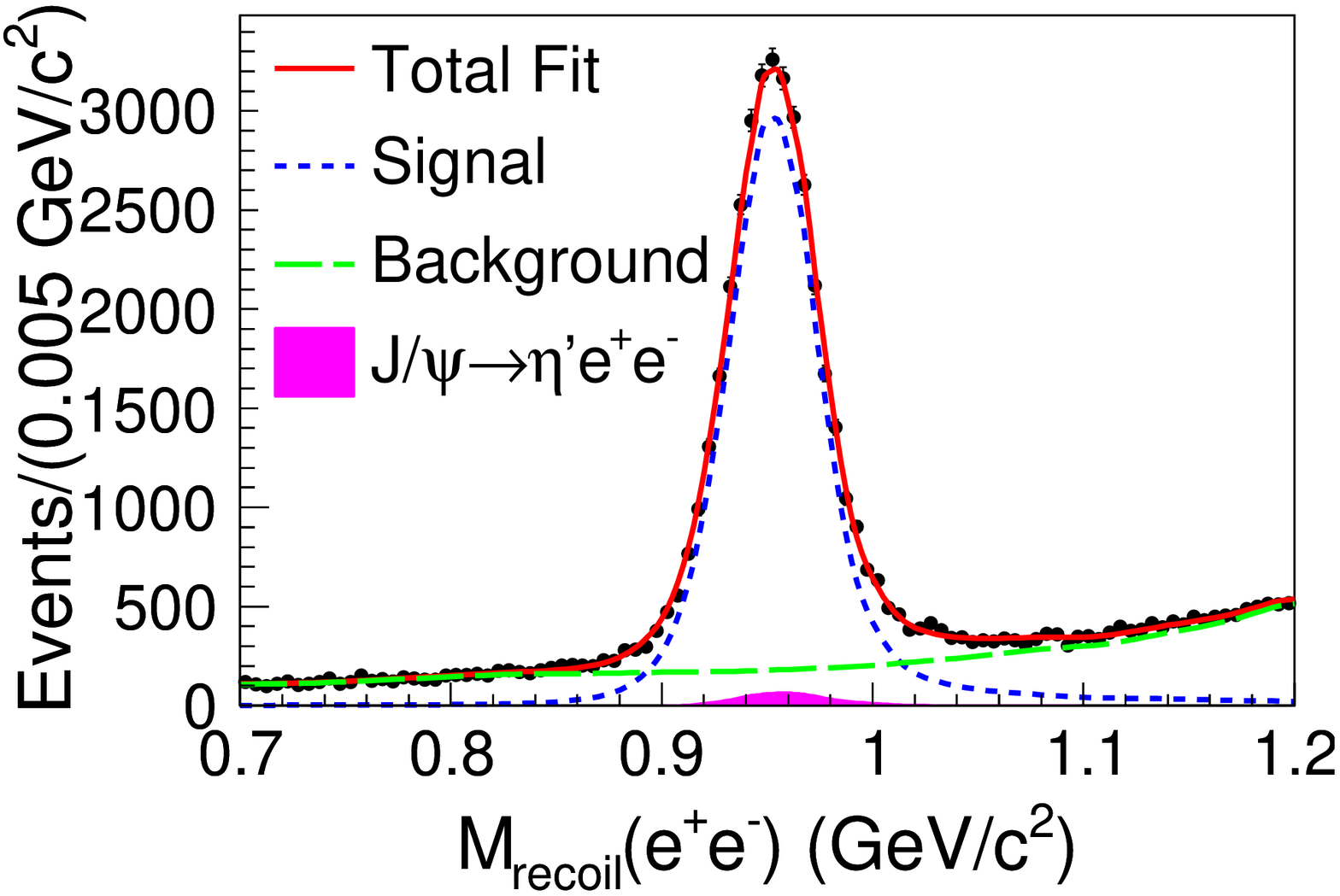}				\put(-30,65){\bf (a)}
		\label{gep_fitresult}
	}
	\end{subfigure}
	\begin{subfigure}
	{
		\includegraphics[width = 0.22\textwidth]{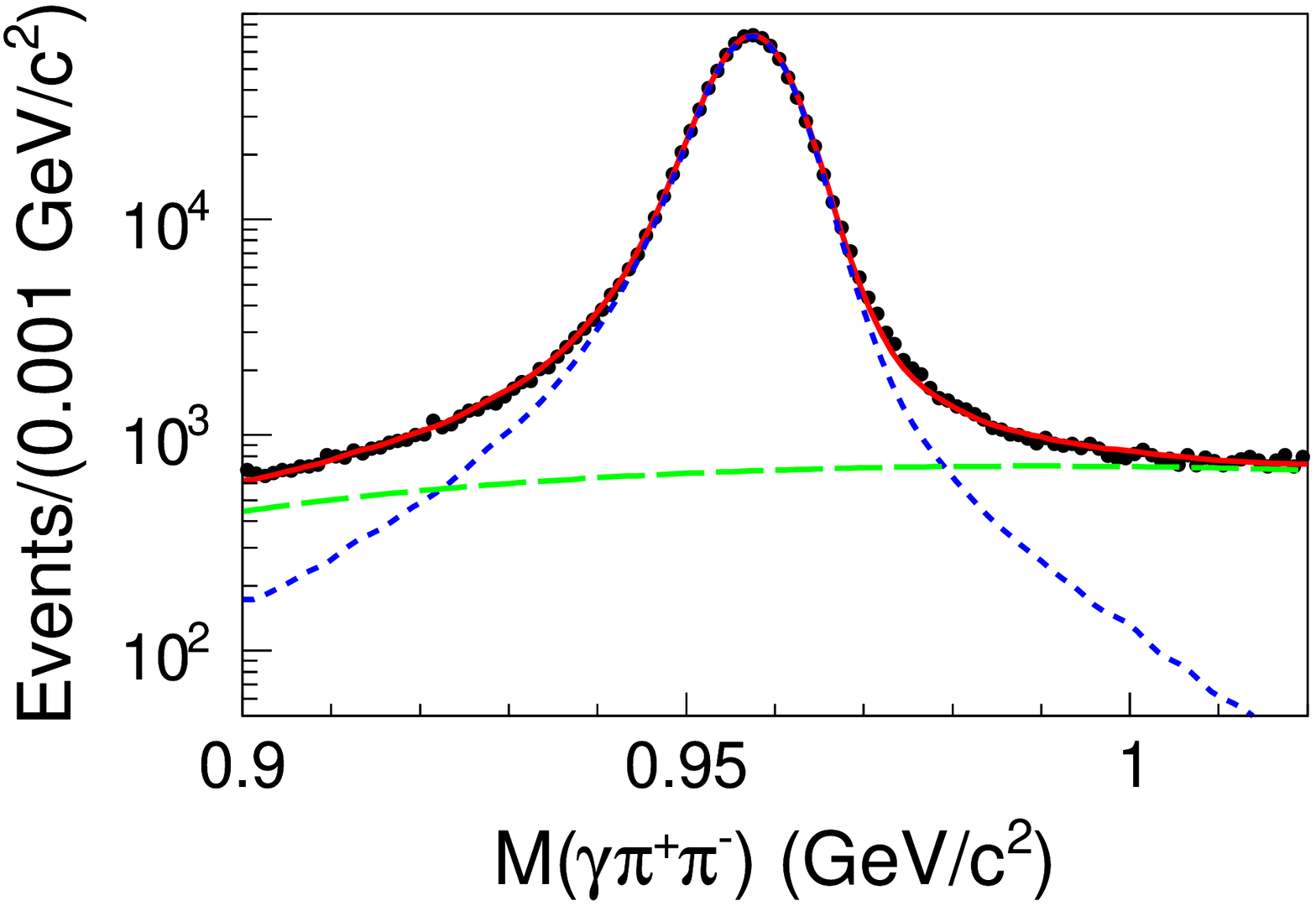}				\put(-30,65){\bf (b)}
		\label{gpipi_fitresult}
	}
	\end{subfigure}
	\begin{subfigure}
	{
		\includegraphics[width = 0.22\textwidth]{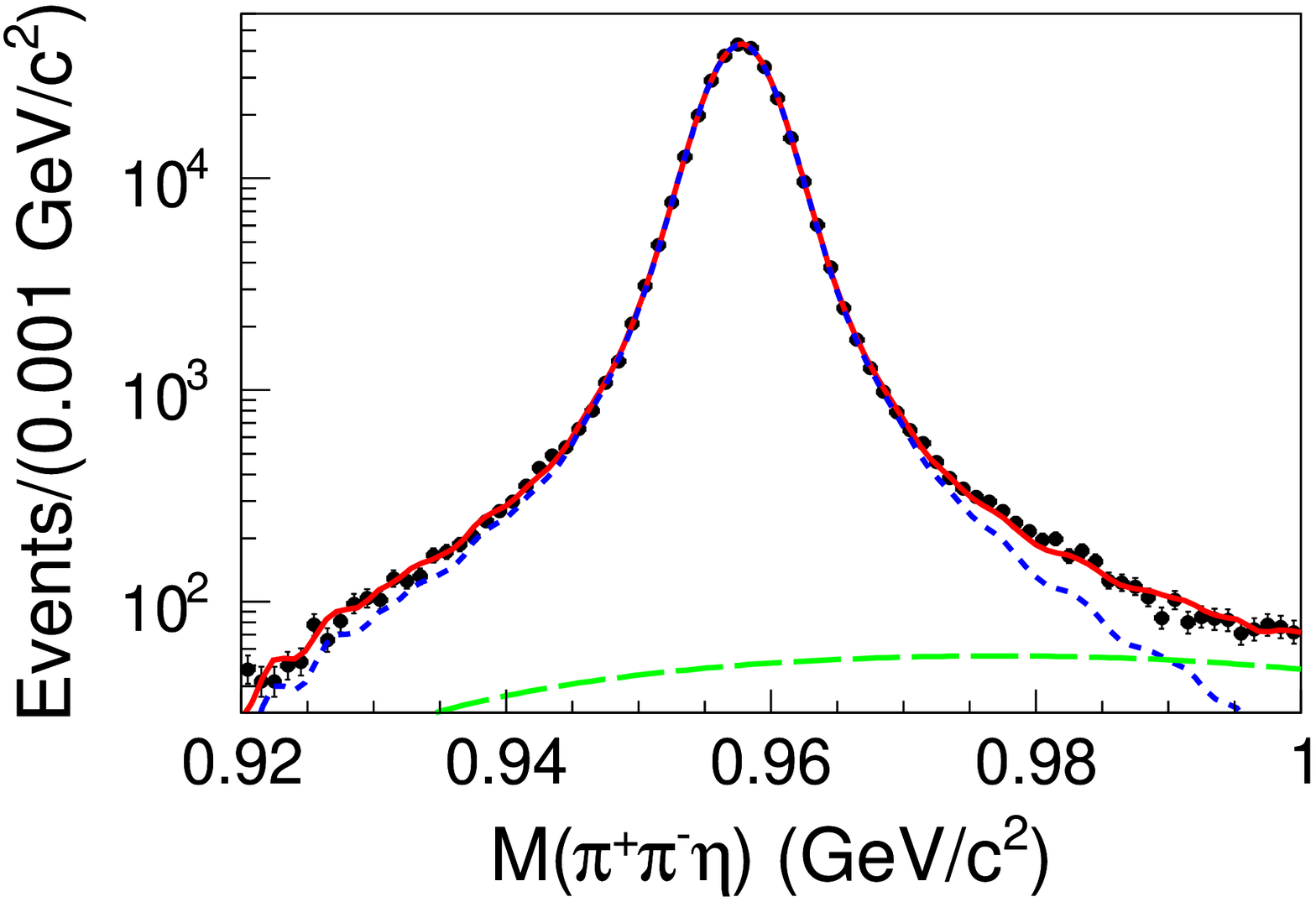}			\put(-30,65){\bf (c)}
		\label{etapipi_fitresult}
	}
	\end{subfigure}
	\begin{subfigure}
	{
		\includegraphics[width = 0.22\textwidth]{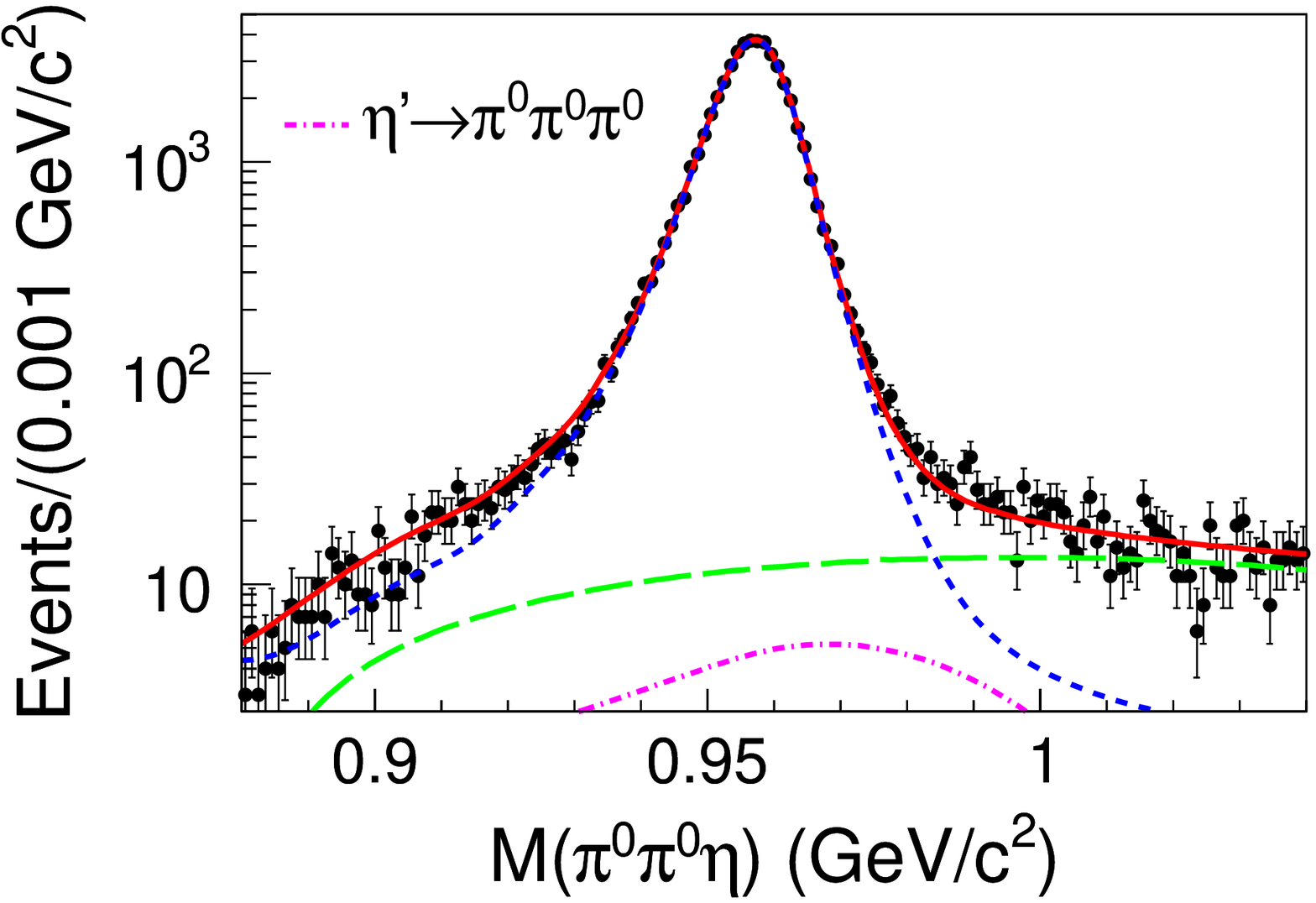}	\put(-30,65){\bf (d)}
		\label{neuetapipi_fitresult_log}
	}
	\end{subfigure}
	\begin{subfigure}
	{
		\includegraphics[width = 0.22\textwidth]{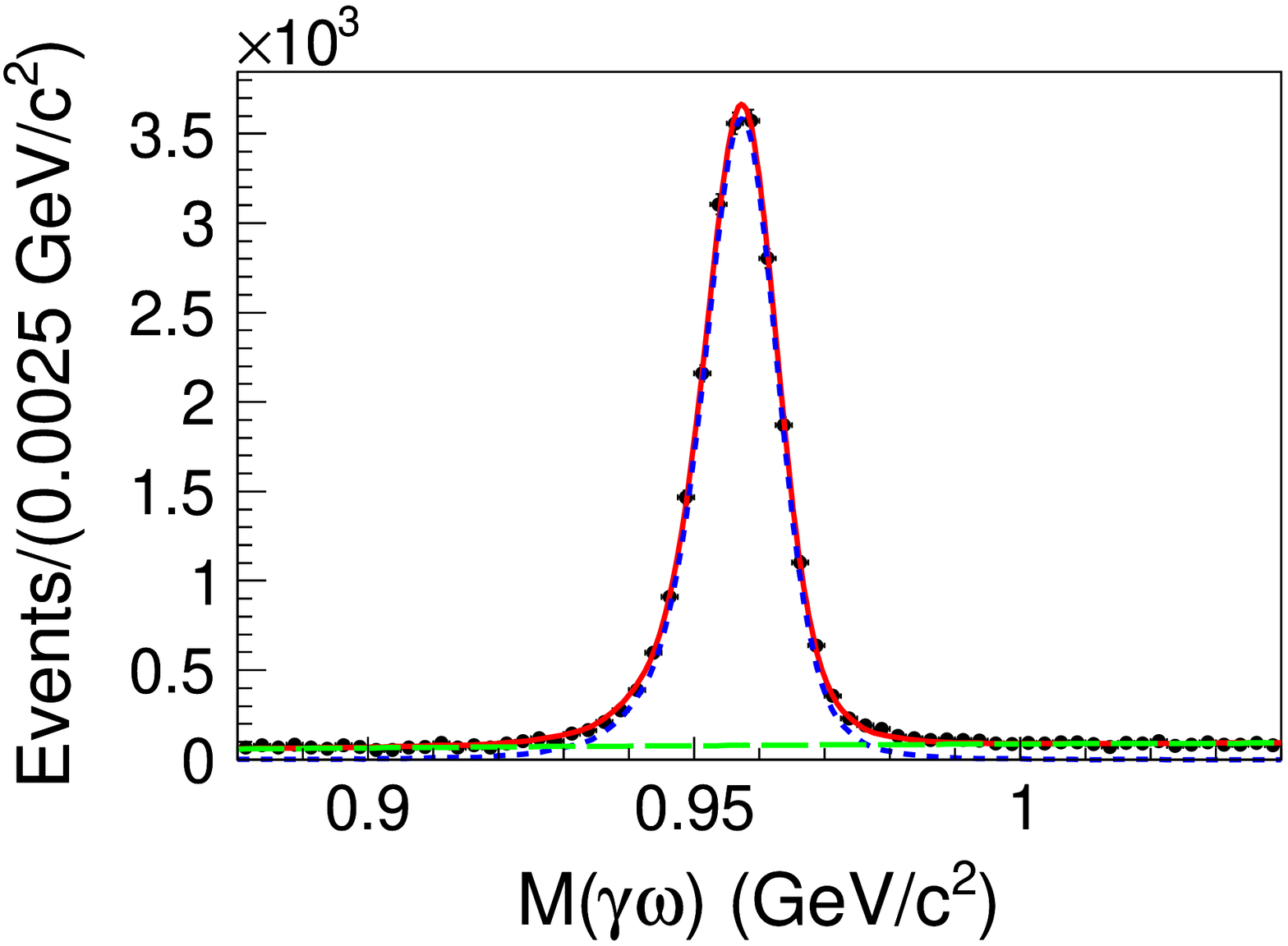}			\put(-30,65){\bf (e)}
		\label{gomega_fitresult}
	}
	\end{subfigure}
	\begin{subfigure}
	{
		\includegraphics[width = 0.22\textwidth]{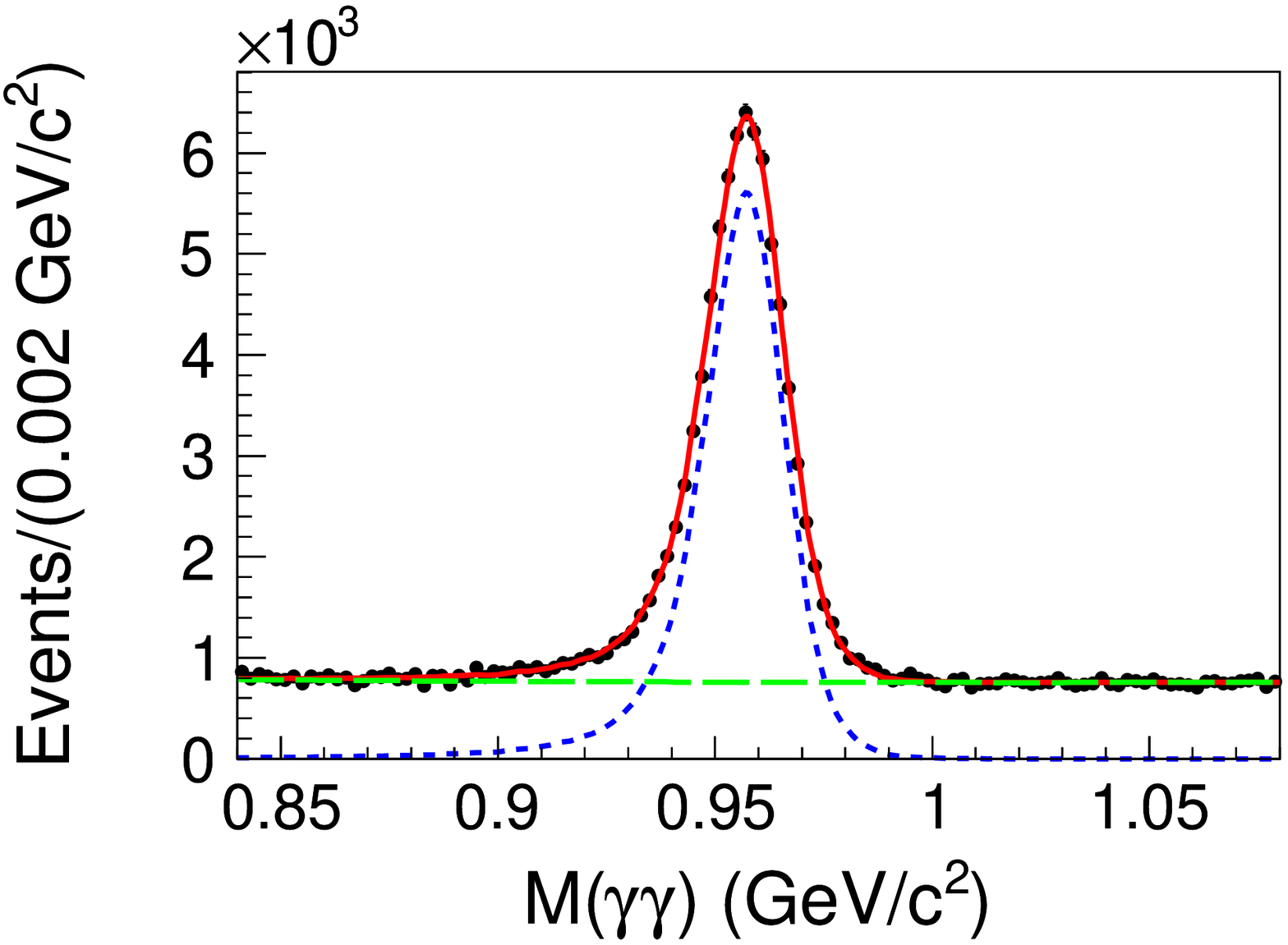}			\put(-30,65){\bf (f)}
		\label{gamgam_fitresult}
	}
	\end{subfigure}
	\caption{Unbinned maximum likelihood fit to the invariant mass spectra. The red solid
	curve shows the result of the fits, the blue dashed line represents the contribution of the signal, and the
	green dashed line represents the smooth background. The pink histogram in (a) is peaking background from
    $\jpsi \ra \etap \epem$, and the pink dashed line in (d) is peaking background from $\etap \ra \pio\pio\pio$.}
	\label{fitresults}
\end{figure}

After the above requirements, the recoil mass spectrum of $e^+e^-$, $M_{\rm recoil} (e^+e^-)$, is shown in Fig.~\ref{gep_fitresult}, where a clear $\eta^\prime$ peak is observed with low background.
To determine the signal yield of the $J/\psi\rightarrow\gamma\eta^\prime$ decays followed by the radiative photon converting into an $e^+e^-$ pair,
 an unbinned extended maximum likelihood fit to  $M_{\rm recoil} (e^+e^-)$  is performed. The probability density function~(PDF) used in the fit consists of three components to describe the mass spectrum:
signal,  peaking background from $J/\psi\rightarrow e^+e^-\eta^\prime$, and combinatorial background. The signal component is modeled by
a MC simulated shape convolved with a Gaussian function to account for the small difference of the mass resolution between MC simulation and data. The parameters of the Gaussian function
are free in the fit. The magnitude and shape of peaking background are obtained from the MC simulation, while the combinatorial background is modeled as the sum
of the background shape obtained from an inclusive MC sample of $1.2 \times 10^9$ $J/\psi$ events, which is  generated with the {\sc LUNDCHARM}  and  {\sc EVTGEN} models~\cite{Ping:2008zz,Lange:2001uf,Chen:2000tv}, and a second-order Chebychev polynomial function, which accounts for
the difference between inclusive MC sample and data.  The fit shown in
Fig.~\ref{gep_fitresult} yields $35980 \pm 234$  $J/\psi\rightarrow\gamma\eta^\prime$ events with the radiative photon converting into an $e^+e^-$ pair.

A MC sample of $J/\psi\rightarrow\gamma\eta^\prime$ in which the $\eta^\prime$ inclusive decays are generated in accordance with the world average  BFs of the established modes. We model $\etap \to \pip\pim\eta$ and  $\etap \to 3\pi$ according to the distributions measured in Refs.~\cite{Ablikim:2017irx, Ablikim:2016frj};  the events of $\etap \to \gamma\pip\pim$, $\pip\pim e^+ e^-$, $\pip\pim\pi^0\pi^0$ and $\pip\pim\pip\pim$ are simulated in accordance with theoretical models~\cite{Wess:1971yu, Witten:1983tw, pipiee, Guo:2011ir}, which have been validated in the previous measurements~\cite{Ablikim:2017fll, Ablikim:2013wfg, Ablikim:2014eoc}; the others, e.g., $\etap \ra \gamma\gamma$ and $\etap \ra \gamma\omega$, are generated with the phase space distribution.
Then the detection efficiency is determined to be $5.15\times 10^{-3}$ according to the MC simulation.
Using this efficiency, we obtained a BF of $\jpsi \to \gamma \etap$ of $(5.27 \pm 0.03) \times 10^{ - 3}$ in which we only present the statistical uncertainty. Moreover, we applied a correction factor $f= \varepsilon_{conv}^{data} / \varepsilon_{conv}^{MC}  =1.0085\pm0.0050$ ~\cite{PEC} to account for the difference in the photon conversion efficiencies.

For the exclusive measurements of $\etap$ decays to $\gamma\pi^+\pi^-$, $\eta\pi^+\pi^-$,
$\eta\pi^0\pi^0$, $\gamma\omega$ and $\gamma\gamma$ with $\pi^0 (\eta)\rightarrow\gamma\gamma$ and $\omega\rightarrow\pi^+\pi^-\pi^0$, the final states are
composed of
$\gamma\gamma\pi^+\pi^-$, $\gamma\gamma\gamma\pi^+\pi^-$, $\gamma\gamma\gamma\gamma\gamma\gamma\gamma$, $\gamma\gamma\gamma\gamma\pi^+\pi^-$ and $\gamma\gamma\gamma$, respectively.
Candidate events are required to satisfy the following common selection criteria.
(1)~Candidate charged tracks and photons are selected with the same method as Ref.~\cite{Ablikim:2017ixv} except that we
only use photons hitting the EMC barrel.
Since $\jpsi \ra \gamma \etap$ is a two-body decay, the radiative
photon from $\jpsi$ decays is mono-energetic with $E=1.4$ GeV,
which makes it easy to distinguish
the photons from $\etap$ decays. The photon with the largest energy is then regarded as
the radiative photon from $\jpsi$. The other photons combined with the charged tracks are used for $\etap$ reconstruction.
(2)~Events must have the correct number of charged tracks with  zero net charge and
at least the minimum number of isolated photons associated with the different final states.
(3)~The selected events are fitted kinematically. The kinematic fit adjusts the track energy and momentum within the measured uncertainties so as to satisfy energy and momentum conservation for the given event hypothesis. This improves the momentum resolution, selects the correct charged-particle assignment for the tracks, and reduces the background. All possible combinations for each signal mode are tested and the combination with the least $\chi^2$ is retained.

In the case of $\etap \ra \gamma \pip \pim$,  a four-constraint (4C)
kinematic fit on the final-state particle candidates is performed and the $\chi_{\rm 4C}^{2}$ is required to be less than $100$.
In order to remove background events with a $\pio$ in the final states, we require that the invariant mass of $\gamma\gamma$
is not in the $\pio$ mass region, $|M_{\gamma \gamma}-m_{\pio}|>0.02$ GeV/$c^{2}$, where $m_{\pio}$
is the nominal mass of the $\pio$~\cite{Olive:2016xmw}.
A MC study of the  $\jpsi$ inclusive decays reveals that the channels  $\jpsi \ra \rho^{0} \pio$ and
$\jpsi \ra \epem(\gamma)$  are the dominant backgrounds, but neither of them produce peaks in the vicinity of the $\etap$ signal in the $\gamma \pip \pim$
invariant-mass spectrum.

For  $\etap \ra \eta \pip \pim$,  a five-constraint (5C) kinematic fit is performed under
 the $\gamma\gamma\gamma\pi^+\pi^-$ hypothesis with the invariant mass of the two photons being constrained to the  $\eta$ mass~\cite{Olive:2016xmw}.
After requiring $\chi_{\rm 5C}^{2}<100$, the remaining data sample contains a very small background level of $0.3$\%,
which is estimated by the events in the $\etap$ mass sideband regions.
By investigating the $\jpsi$ inclusive MC sample, the dominant background contributions are found to be
from $\jpsi \to \gamma \eta \pip \pim$ and
$\jpsi \to \gamma \gamma\rho$, but no peaking background
appears in the  $\eta \pip \pim$ invariant mass distribution around the $\etap$ signal region.

To detect $\etap \ra \eta \pio \pio$,  one-constraint (1C) kinematic fits are performed on the $\pi^0$ ($\eta$) candidates reconstructed from photon pairs with the  invariant mass of the two photons being constrained to the
$\pi^0$ ($\eta$) mass, and $\chi^2_{\rm 1C}$ is required to be less than 25.   Then a seven-constraint (7C)
kinematic fit (two $\pi^0$ and one $\eta$ mass are also constrained in addition to the four energy-momentum constraints) is performed  under the hypothesis of $J/\psi\rightarrow\gamma\pi^0\pi^0\eta$ and
$\chi_{\rm 7C}^{2}<100$ is required.
After that the candidate events, as illustrated by the mass spectrum of $\eta\pi^0\pi^0$ in Fig.~\ref{neuetapipi_fitresult_log}, are almost background free. A MC study shows that the background events of  $\jpsi \ra \gamma \etap, \etap \ra \pio \pio \pio$ contribute to a small peak in the $ \eta \pio \pio$ mass distribution around the $\etap$ signal region, which is considered
in the signal extraction.

To select  $\etap \ra \gamma \omega$ candidates,  five-constraint (5C) kinematic fits
are performed with the invariant mass of all combinations of any two photons being constrained to the $\pio$ mass,
and $\chi^2_{\rm 5C}$ is required to be less than 50. We require the $\pip \pim \pio$ invariant
mass is in the $\omega$ signal region, $|M_{\pip \pim \pio}-m_{\omega}|<0.03$ GeV/$c^{2}$, where $m_{\omega}$ is the nominal mass of
the $\omega$~\cite{Olive:2016xmw}. If the recoil mass of the $\omega$ satisfies $|M_{\omega}^{\rm rec}-m_{\pio}|<0.025$ GeV/$c^{2}$ or
$|M_{\omega}^{\rm rec}-m_{\eta}|<0.035$ GeV/$c^{2}$, the events are rejected to  suppress background contributions from $\jpsi \ra \omega \eta$ and $\jpsi \ra \omega \pio$.
According to a MC study using the $J/\psi$ inclusive sample, the remaining background events
mainly come from $\jpsi \ra b_1(1235)^0 \pio$ with $b_1(1235)^0 \ra \omega \pio$
and $\jpsi \ra \omega \pio \pio$,  but neither of them produces a peak in the $\gamma\omega$ mass spectrum near the $\eta^\prime$ mass.

For the decay of  $\etap \ra \gamma \gamma$, a 4C-kinematic fit is applied, and events with $\chi_{\rm 4C}^{2} < 60$ are selected.  Since there is a small probability that the energy of one photon from the $\eta^\prime$ decay is larger than
that of the radiative photon,  the mass distributions of the three  photon pairs for each event are plotted in Fig.~\ref{gamgam_fitresult}, where an  $\eta^\prime$ signal is clearly observed
above a smooth background due to wrong   $\gamma\gamma$ combinations plus other background sources.

After applying the above requirements, the mass spectra of $\gamma\pi^+\pi^-$, $\eta\pi^+\pi^-$,
$\eta\pi^0\pi^0$, $\gamma\omega$ and $\gamma\gamma$ are shown in Figs.~\ref{fitresults}(b)-(f), where the $\eta^\prime$ signals for different exclusive decays
are clearly observed. The corresponding signal yields are obtained by performing the extended unbinned maximum likelihood fits to the above mass spectra.
The PDF function consists of a signal and various background contributions. The signal component is modeled as the MC simulated signal shape convolved with a Gaussian function to
account for the difference in the mass resolution between data and MC simulation. The considered background components are subdivided into two classes: (i)~the non-peaking background, which is described
with a first-order or second-order Chebychev polynomial function; (ii)~the peaking background in $\etap \to \eta \pi^0 \pi^0$, $e.g.$, $\jpsi \ra \gamma \etap$, $\etap \ra \pio\pio\pio$,
,  which is described by the shape determined via a MC simulation and the corresponding magnitude is estimated according to the corresponding branching fraction from PDG~\cite{Olive:2016xmw}. The fit results for the signal yields are listed in Table~\ref{tab:sum} and the projections of the fit on the mass spectra for different exclusive decays are shown in Figs.~\ref{fitresults}(b)-(f), respectively.

According to Eq.~(\ref{eq:gamgambf}), the  BFs for these five dominant decays of $\eta^\prime$  are presented in Table~\ref{tab:sum}, where the first uncertainties are statistical and the second systematic.

\begin{table*}[htbp]
\begin{center}
\caption{Summary of the measured BFs for $\etap$ decays. $N_{\etap \to X }^{\rm obs}$ is the signal yield from the fits, $\varepsilon _{\etap \to X }$ is the
detection efficiency, and $\mathcal{B}$ is the determined BF.}
\label{tab:sum}
\begin{tabular}{l  c  c  c  c c c }\hline\hline
Decay Mode								&$N_{\etap \to X }^{\rm obs}$	&$\varepsilon _{\etap \to X }$(\%)	& \multicolumn{2}{c}{$\mathcal{B}(\etap \to X)$(\%)} &\multicolumn{2}{c}{$\mathcal{B}/\mathcal{B}(\eta ' \to \eta \pi^{+} \pi^{-})$}\\
& & &This measurement &PDG~\cite{Olive:2016xmw} &This measurement	&CLEO \cite{Pedlar:2009aa} \\ \hline
$\eta ' \to \gamma \pi^{+} \pi^{-}$		&913106$\pm$1052						&44.11				&29.90$\pm$0.03$\pm$0.55		&	28.9$\pm$0.5\phantom{1}&0.725$\pm$0.002$\pm$0.010\phantom{1}			&0.677$\pm$0.024$\pm$0.011\\
$\eta ' \to \eta \pi^{+} \pi^{-}$		&312275$\pm$570\phantom{1}				&27.75				&41.24$\pm$0.08$\pm$1.24		&	42.6$\pm$0.7\phantom{1} & ...  & ...\\
$\eta ' \to \eta \pi^{0} \pi^{0}$		&\phantom{1}51680$\pm$238\phantom{1}	&\phantom{1}9.08	&21.36$\pm$0.10$\pm$0.92		&	22.8$\pm$0.8\phantom{1}&0.518$\pm$0.003$\pm$0.021\phantom{1}			&0.555$\pm$0.043$\pm$0.013\\
$\eta ' \to \gamma \omega$				&\phantom{1}22749$\pm$163\phantom{1}	&14.98	 		&\phantom{1}\phantom{1}2.489$\pm$0.018$\pm$0.074	&2.62$\pm$0.13 &0.0604$\pm$0.0005$\pm$0.0012			&0.055$\pm$0.007$\pm$0.001\\
$\eta ' \to \gamma \gamma$          	&\phantom{1}70669$\pm$349\phantom{1}	&43.79	    	&\phantom{1}\phantom{1}2.331$\pm$0.012$\pm$0.035	& 2.22$\pm$0.08 	&0.0565$\pm$0.0003$\pm$0.0015			&0.053$\pm$0.004$\pm$0.001\\

\hline \hline
\end{tabular}
\end{center}
\end{table*}

Sources of systematic uncertainties for the BF measurements for $\etap$ decays can be divided into two categories:
those from the $\etap$ exclusive measurements and those from the inclusive measurement.

Systematic uncertainties from the $\etap$ exclusive measurements are mainly from the MDC tracking efficiency, the photon detection efficiency,
the kinematic fit, and the fit procedure.
The MDC tracking efficiency for the charged pion is studied with a control sample of $\jpsi \ra \rho \pi$,  and
the weighted average uncertainties are obtained using bins of transverse momentum~\cite{Ablikim:2017fll}.
The systematic uncertainty due to the photon detection efficiency
is studied with a control sample of $\jpsi \ra \pip \pim \pio$~\cite{Ablikim:2015umt}.
In $\jpsi \ra \gamma \etap$, the radiative
photon carries a unique energy of $1.4$ GeV. The detection efficiency of the radiative photon  is studied with $\jpsi \ra \gamma \etap, \etap \to \gamma \pip \pim$.
For the uncertainties in the reconstruction of the $\eta$ and $\pio$, we use the result of a study described in Ref.~\cite{Ablikim:2010zn}.
The uncertainty associated with the kinematic fit arises from the inconsistency between the data and the MC simulation.
For decay processes including charged tracks in the final states and decay processes with
purely neutral particles in the final states, the uncertainties are estimated with helix parameter
correction~\cite{Ablikim:2012pg}  and photon energy correction~\cite{Ablikim:2016exh}, respectively.
The sources of systematic uncertainty in the fit procedures are estimated by varying the fit ranges, background shapes and signal shapes in each fit, uncertainty form peaking background in
$\etap \to \eta \pi^0 \pi^0$ is negligible.
To estimate the systematic uncertainty due to the kinematics of the $\etap$
three-body decays, we generate the $\eta ' \to \gamma \pi^{+} \pi^{-}$, $\eta ' \to \eta \pi^{+} \pi^{-}$ and $\eta ' \to \eta \pi^{0} \pi^{0}$
signal MC samples with parameters from different measurements~\cite{Dorofeev:2006fb,Blik:2009zz,Ablikim:2017irx}. The changes in the reconstruction efficiency are taken as the
systematic uncertainties.

In addition to the above exclusive systematic sources, the uncertainty from the $\etap$ inclusive measurement is included in the measurement of the BFs.  Note that the efficiencies of the electron tracking and the photon conversion
reconstruction criteria cancel in the photon conversion efficiency correction.
Thus the uncertainties on the $\etap$ inclusive measurement consist of uncertainties in the fit procedure, the number of peaking
background events from $\jpsi \ra \epem \etap $, the statistical uncertainty on $N^{\rm obs}_{J/\psi\rightarrow\gamma\eta^\prime}$ and the uncertainty in the correction factor applied to the photon-conversion efficiency.
The total systematic uncertainty from the $\etap$ inclusive measurement is 0.9\% and it is indicated as the $\etap$ inclusive
uncertainty in Table \ref{tab:sysunsum}.

In the measurement of the BF for $\jpsi \ra \gamma \etap$, the sources of systematic uncertainty are the same as those for the $\etap$ inclusive measurement except
that the uncertainty of the number of $\jpsi$ decays \cite{Ablikim:2016fal} is included instead of the statistical uncertainty of $N^{\rm obs}_{J/\psi\rightarrow\gamma\eta^\prime}$.

Table \ref{tab:sysunsum} summarizes all contributions to the systematic uncertainties on the BF measurements.
In each case, the total systematic uncertainty is given by the quadratic sum of the individual
contributions, assuming all sources to be independent.

\begin{table}[htbp]
\begin{center}
\caption{Summary of all sources of systematic uncertainties (in \%) in the $\etap$ and $\jpsi$ BF measurements. The ellipsis "..." indicates
that the uncertainty is not applicable. \uppercase\expandafter{\romannumeral1}-\uppercase\expandafter{\romannumeral5} represent $\etap \to$
$\gamma\pi^+\pi^-$, $\eta\pi^+\pi^-$,
$\eta\pi^0\pi^0$, $\gamma\omega$ and $\gamma\gamma$, respectively, while \uppercase\expandafter{\romannumeral6} represents $\jpsi \to \gamma \etap$. }
\label{tab:sysunsum}
\begin{tabular}{ l  p{0.8cm}<{\centering}  p{0.8cm}<{\centering}  p{0.8cm}<{\centering}  p{0.8cm}<{\centering}  p{0.8cm}<{\centering} p{0.8cm}<{\centering} }\hline\hline
Sources   &  \uppercase\expandafter{\romannumeral1} & \uppercase\expandafter{\romannumeral2}
&\uppercase\expandafter{\romannumeral3} & \uppercase\expandafter{\romannumeral4}
&\uppercase\expandafter{\romannumeral5} & \uppercase\expandafter{\romannumeral6} \\ \hline

Tracking                    &1.3	&2.3	&...	&1.9	&...    &...	\\
Radiative $\gamma$          &0.2	&0.2	&0.2	&0.2	&0.2	&...   \\
$\gamma$ detection          &0.5	&1.0	&3.0	&1.5	&1.0	&...   \\
$\pi^{0}$ reconstruction    &...	&...	&2.0	&1.0	&...	&...   \\
$\eta$ reconstruction       &...	&1.0	&1.0	&...	&...	&...   \\
Kinematics fit              &0.1	&0.1	&1.7	&0.5	&0.5	&...   \\
Fit range                   &0.2	&0.2	&0.2	&0.1	&0.2	&0.3   \\
Signal shape                &0.2    &0.1    &0.1    &0.3    &0.1    &0.2    \\
Background shape            &0.3	&0.4	&0.1	&0.1	&0.2	&0.2  \\
Peaking background          &...	&...	&...	&...	&...	&0.2  \\
Physical model              &0.6	&0.7	&0.5	&...	&...	&...  \\
BFs                         &...	&0.5	&0.5	&0.8	&...	&...   \\
$f$                         &...	&...	&...	&...	&...    &0.5    \\
$\etap$ inclusive           &0.9	&0.9	&0.9	&0.9	&0.9    &...    \\
$N_{\jpsi}$                 &...	&...	&...	&...	&...    &0.53    \\  \hline
Total                       &1.8	&3.0	&4.3	&3.0&1.5	&0.9   \\
\hline \hline
\end{tabular}
\end{center}
\end{table}


In summary, using a  data sample of $(1310.6 \pm 7.0) \times 10^6~\jpsi$ events collected  with the BESIII detector,  we present
a model-independent measurement of the BF for $J/\psi\rightarrow\gamma\eta^\prime$ by analyzing events where the radiative photon converts into an $e^+e^-$ pair.
The BF of  $J/\psi\rightarrow\gamma\eta^\prime$ is determined to be $(5.27\pm0.03\pm0.05)\times 10^{-3}$, which is in agreement with the world average value~\cite{Olive:2016xmw}, but with a significantly improved precision.
Taking advantage of the sample of $\eta^\prime$ inclusive decays tagged by
$J/\psi\rightarrow\gamma\eta^\prime$ events with photon conversion,  the absolute BFs of five dominant decays of the $\etap$ are presented in Table~\ref{tab:sum} and are measured independently for the first time, which are in agreement with the PDG values~\cite{Olive:2016xmw}.
In addition, we give the relative BFs for $\etap$ decays as presented in Table~\ref{tab:sum},
which are in agreement with CLEO's result~\cite{Pedlar:2009aa} within two standard deviations.
The precision of our measurements is a factor 2 to 4 better than that of CLEO.
The comparisons of the decay widths of $\eta ' \to \eta \pip \pim$ and $\eta ' \to \eta \pi^{0} \pi^{0}$ with
different theoretical approaches, including the chiral unitary approach~\cite{th1}, the chiral perturbation theory~\cite{th5} and the chiral effective field theory~\cite{th4}, are presented in Table~\ref{tab:thr}. Here the measured decay widths are obtained using the
$\etap$ total decay width $\Gamma(\etap) = 0.196 \pm 0.009$~MeV~\cite{Olive:2016xmw}. Our results are in good agreement with
the theoretical estimation.
The photon conversion method in this Letter can also be applied in other measurements using $\jpsi$
radiative decays, such as the decay $J/\psi \to \gamma\eta$.

\begin{table}[htbp]
\begin{center}
\begin{small}
\caption{Comparison of measured decay widths~(keV) with theoretical calculations.}
\label{tab:thr}
\begin{tabular}{ l  p{2.5cm}<{\centering}  p{2.5cm}<{\centering} }\hline\hline
        &$\Gamma(\eta ' \to \eta \pi^{+} \pi^{-})$	&$\Gamma(\eta ' \to \eta \pi^{0} \pi^{0})$	\\	\hline
Reference~\cite{th1}	&77.7			&43.8  \\
Reference~\cite{th5}	&83.6$\pm$0.8			&42.9$\pm$0.3		\\
Reference~\cite{th4}	&81$\pm$4			&46$\pm$3		\\
This measurement	&80.8$\pm$4.4			&41.8$\pm$2.6		\\
\hline \hline
\end{tabular}
\end{small}
\end{center}
\end{table}

\input{acknowledgement.tex}


\end{document}

%% file: authors.tex
\author
{
\begin{small}
\begin{center}
M.~Ablikim$^{1}$, M.~N.~Achasov$^{10,d}$, S. ~Ahmed$^{15}$, M.~Albrecht$^{4}$, M.~Alekseev$^{55A,55C}$, A.~Amoroso$^{55A,55C}$, F.~F.~An$^{1}$, Q.~An$^{52,42}$, Y.~Bai$^{41}$, O.~Bakina$^{27}$
% ...
\\
\vspace{0.2cm}
(BESIII Collaboration)\\
\vspace{0.2cm} {\it
$^{1}$ Institute of High Energy Physics, Beijing 100049, People's Republic of China\\
% ...
$^{4}$ Bochum Ruhr-University, D-44780 Bochum, Germany\\
% ...
$^{10}$ G.I. Budker Institute of Nuclear Physics SB RAS (BINP), Novosibirsk 630090, Russia\\
% ...
$^{15}$ Helmholtz Institute Mainz, Johann-Joachim-Becher-Weg 45, D-55099 Mainz, Germany\\
% ...
$^{27}$ Joint Institute for Nuclear Research, 141980 Dubna, Moscow region, Russia\\
% ...
$^{41}$ Southeast University, Nanjing 211100, People's Republic of China\\
$^{42}$ State Key Laboratory of Particle Detection and Electronics, Beijing 100049, Hefei 230026, People's Republic of China\\
% ...
$^{52}$ University of Science and Technology of China, Hefei 230026, People's Republic of China\\
% ...
$^{55}$ (A)University of Turin, I-10125, Turin, Italy; (B)University of Eastern Piedmont, I-15121, Alessandria, Italy; (C)INFN, I-10125, Turin, Italy\\
% ...
$^{d}$ Also at the Novosibirsk State University, Novosibirsk, 630090, Russia\\
% ...
}
\end{center}
\vspace{0.4cm}
\end{small}
}

%% file: acknowledgement.tex
The BESIII collaboration thanks the staff of BEPCII and the IHEP computing center for their strong support. This work is supported in part by National Key Basic Research Program of China under Contract No. 2015CB856700; National Natural Science Foundation of China (NSFC) under Contracts Nos. 11335008, 11425524, 11625523, 11635010, 11675184, 11735014; the Chinese Academy of Sciences (CAS) Large-Scale Scientific Facility Program; the CAS Center for Excellence in Particle Physics (CCEPP); Joint Large-Scale Scientific Facility Funds of the NSFC and CAS under Contracts Nos. U1532257, U1532258, U1732263; CAS Key Research Program of Frontier Sciences under Contracts Nos. QYZDJ-SSW-SLH003, QYZDJ-SSW-SLH040; 100 Talents Program of CAS; INPAC and Shanghai Key Laboratory for Particle Physics and Cosmology; Shandong Natural Science Funds for Distinguished Young Scholar under Contract No. JQ201402; German Research Foundation DFG under Contracts Nos. Collaborative Research Center CRC 1044, FOR 2359; Istituto Nazionale di Fisica Nucleare, Italy; Koninklijke Nederlandse Akademie van Wetenschappen (KNAW) under Contract No. 530-4CDP03; Ministry of Development of Turkey under Contract No. DPT2006K-120470; National Science and Technology fund; The Swedish Research Council; U. S. Department of Energy under Contracts Nos. DE-FG02-05ER41374, DE-SC-0010118, DE-SC-0010504, DE-SC-0012069; University of Groningen (RuG) and the Helmholtzzentrum fuer Schwerionenforschung GmbH (GSI), Darmstadt.

%% file: bf_etaprime_0319.bbl
\begin{thebibliography}{99}
\bibitem{PEC}
The photon conversion efficiency correction factor is obtained by investigating the photon conversion efficiency in the process  $e^+e^-\rightarrow\gamma\gamma$ using data samples taken at center-of-mass energies between 2.000 GeV and 3.773 GeV.
The photon conversion efficiency is defined as $\varepsilon_{conv} = N_{conv}/(2 \cdot N_{\gamma\gamma})$, where $N_{conv}$ is the number of observed $\epem \to \gamma \gamma$ events
with one converted photon while $N_{\gamma\gamma}$ is the number of observed $\epem \to \gamma \gamma$ events.
\end{thebibliography}
